# Fermi Level Controlled Ultrafast Demagnetization Mechanism in Half-Metallic Heusler Alloy

Santanu Pan[1], Takeshi Seki[2,3], Koki Takanashi[2,3,4], and Anjan Barman[1,*]

[1]Department of Condensed Matter Physics and Material Sciences, S. N. Bose National Centre for Basic Sciences, Block JD, Sector III, Salt Lake, Kolkata 700 106, India.
[2]Institute for Materials Research, Tohoku University, Sendai 980-8577, Japan.
[3]Center for Spintronics Research Network, Tohoku University, Sendai 980-8577, Japan.
[4]Center for Science and Innovation in Spintronics, Core Research Cluster, Tohoku University, Sendai 980-8577, Japan.
[*]E-mail: abarman@bose.res.in

The electronic band structure-controlled ultrafast demagnetization mechanism in $Co_2Fe_xMn_{1-x}Si$ Heusler alloy is underpinned by systematic variation of composition. We find the spin-flip scattering rate controlled by spin density of states at Fermi level is responsible for non-monotonic variation of ultrafast demagnetization time ($\tau_M$) with $x$ with a maximum at $x = 0.4$. Furthermore, Gilbert damping constant exhibits an inverse relationship with $\tau_M$ due to the dominance of inter-band scattering mechanism. This establishes a unified mechanism of ultrafast spin dynamics based on Fermi level position.

The tremendous application potential of spin-polarized Heusler alloys in advanced spintronics devices ignites immense interest to investigate the degree and sustainability of their spin-polarization under various conditions [1-4]. However, interpreting spin-polarization from the conventional methods such as photoemission, spin transport measurement, point contact Andreev reflection and spin-resolved positron annihilation are non-trivial [5-7]. In the quest of developing alternative methods, Zhang *et al*. demonstrated that all-optical ultrafast demagnetization measurement is a reliable technique for probing spin-polarization [8]. They observed a very large ultrafast demagnetization time as a signature of high spin-polarization in half-metallic $CrO_2$. However, Co-based half-metallic Heusler alloys exhibit a comparatively smaller ultrafast demagnetization time (~ 0.3 ps) which raised a serious debate on the perception of ultrafast demagnetization mechanism in Heusler alloys [9-11]. A smaller demagnetization time in Heusler alloys than in $CrO_2$ is explained due to the smaller effective band gap in the minority spin band and enhanced spin-flip scattering (SFS) rate [9]. However, further experimental evidence shows that the amount of band gap in minority spin band cannot be the only deciding factors for SFS mediated ultrafast demagnetization efficiency [10]. Rather, one also has to consider the efficiency of optical excitation for majority and minority spin bands as well as the optical pump-induced hole dynamics below Fermi energy ($E_F$). Consequently, a clear interpretation of spin-polarization from ultrafast demagnetization measurement requires a clear and thorough understanding of its underlying mechanism. Since its inception in 1996 [12], several theoretical models and experimental evidences based on different microscopic mechanisms, e.g. spin-flip scattering (SFS) and super-diffusive spin current have been put forward to interpret ultrafast demagnetization [13-20]. However, the preceding proposals are complex and deterring to each other. This complexity increases even more in case of special class of material such as the Heusler alloys. The electronic band structure and the associated position of Fermi level can be greatly tuned by tuning the alloy composition of Heusler alloy [21,22]. By utilizing this tunability, here, we experimentally demonstrate that the ultrafast demagnetization mechanism relies on the spin density of states at Fermi level in case of half-metallic Heusler alloy system. We extracted the value of ultrafast demagnetization time using three temperature modelling [23] and found its non-monotonic dependency on alloy composition (*x*). We have further showed that the Gilbert damping and ultrafast demagnetization time are inversely proportional in CFMS Heusler alloys suggesting the inter-band scattering as the primary mechanism behind the Gilbert damping in CFMS Heusler alloys. Our work has established a unified theory of ultrafast spin dynamics.

A series of Co$_2$Fe$_x$Mn$_{1-x}$Si (CFMS) thin films have been deposited using magnetron co-sputtering system for our investigation with $x$ = 0.00, 0.25, 0.40, 0.50, 0.60, 0.75 and 1.00. The thickness of the CFMS layer was fixed at 30 nm. It is imperative to study the crystalline phase which is the most crucial parameter that determines other magnetic properties of Heusler alloy. Prior to the magnetization dynamics measurement, we investigate both the crystalline phase as well as growth quality of all the samples. Fig. 1A shows the *ex-situ* x-ray diffraction (XRD) pattern for all the samples. The well-defined diffraction peak of CFMS (400) at $2\theta$ = 66.50º indicates that the samples are well crystalline having cubic symmetry. The intense superlattice peak at $2\theta$ = 31.90º represents the formation of B2 phase. The presence of other crucial planes are investigated by tilting the sample $x$ = 0.4 by 54.5º and 45.2º from the film plane to the normal direction, respectively and observed the presence of (111) superlattice peak along with the (220) fundamental peak as shown in Fig. 1B and 1C. The presence of (111) superlattice peak confirms the best atomic site ordering in the desired L2$_1$ ordered phase, whereas the (220) fundamental peak results from the cubic symmetry. The intensity ratios of the XRD peaks are analysed to obtain the microscopic atomic site ordering which remain same for the whole range of $x$ (given in Supplemental Materials). The epitaxial growth of the thin films is ensured by observing the *in-situ* reflection high-energy electron diffraction (RHEED) images. The square shaped hysteresis loops obtained using in-plane bias magnetic field shows the samples have in-plane magnetization. The nearly increasing trend of saturation magnetization with alloy composition ($x$) follow the Slater-Pauling curve. In-depth details of sample deposition procedure, RHEED pattern and the hysteresis loops are provided in the Supplemental Materials [24]. The ultrafast demagnetization dynamics measurements using time-resolved magneto-optical Kerr effect (TRMOKE) magnetometer have been performed at a fixed probe fluence of 0.5 mJ/cm$^2$, while the pump fluence have been varied over a large range. Details of the TRMOKE technique is provided in Supplemental Materials [24]. The experimental data of variation of Kerr rotation corresponding to the ultrafast demagnetization measured for pump fluence = 9.5 mJ/cm$^2$ is plotted in Fig. 2A for different values of $x$. The data points are then fitted with a phenomenological expression derived from the three temperature model-based coupled rate equations in order to extract the ultrafast demagnetization time ($\tau_M$) and fast relaxation ($\tau_E$) time [23], which is given below:

$$-\Delta\theta_k = \{[\frac{A_1}{(t/t_0+1)^{1/2}} - \frac{(A_2\tau_E - A_1\tau_M)}{(\tau_E - \tau_M)}e^{-t/\tau_M} - \frac{\tau_E(A_1 - A_2)}{(\tau_E - \tau_M)}e^{-t/\tau_E}]H(t) + A_3\delta(t)\} \otimes G(t) \quad (1)$$

where $A_1$ represents the magnetization amplitude after equilibrium between electron, spin and lattice is restored, $A_2$ is proportional to the maximum rise in the electron temperature and $A_3$ represents the state filling effects during pump-probe temporal overlap described by a Dirac delta function. $H(t)$ and $\delta(t)$ are the Heaviside step and Dirac delta functions, and $G(t)$ is a Gaussian function which corresponds to the laser pulse.

The $\tau_M$ extracted from the fits are plotted as a function of $x$ in Fig. 2B, which shows a slight initial increment followed by a sharp decrement with $x$. In addition, the ultrafast demagnetization rate is found to be slower in the present Heusler alloys than in the 3d metals [9]. The theoretical calculation of electronic band structure of CFMS showed no discernible change in the amount of energy gap in minority spin band but a change in position of $E_F$ with $x$, which lies at the two extreme ends of the gap for $x = 0$ and $x = 1$. Thus, the variation of $\tau_M$ with $x$ clearly indicates that the composition dependent $E_F$ position is somehow responsible for the variation in $\tau_M$. This warrants the investigation of ultrafast demagnetization with continuously varying $x$ values between 0 and 1. However, a majority of earlier investigations [10,11,25], being focused on exploring the ultrafast demagnetization only of $Co_2MnSi$ ($x = 0$) and $Co_2FeSi$ ($x = 1$), lack a convincing conclusion about the role of electronic band structure on ultrafast demagnetization mechanism.

In case of 3d transition metal ferromagnets, Elliott-Yafet (EY)-based SFS mechanism is believed to be responsible for rapid rise in the spin temperature and ultrafast demagnetization [15]. In this theory it has been shown that a scattering event of an excited electron with a phonon changes the probability to find that electron in one of the spin states, namely the majority spin-up ($\uparrow$) or minority spin-down ($\downarrow$) state, thereby delivering angular momentum to the lattice from the electronic system. It arises from the band mixing of majority and minority spin states with similar energy value near the Fermi surface owing to the spin-orbit coupling (SOC). The spin mixing parameter ($b^2$) from the EY theory [26,27] is given by:

$$\langle b^2 \rangle = \min \overline{(\langle \psi_k | \uparrow \rangle \langle \uparrow | \psi_k \rangle, \langle \psi_k | \downarrow \rangle \langle \downarrow | \psi_k \rangle)} \qquad (2)$$

where $\psi_k$ represent the eigen-state of a single electron and the bar denotes a defined average over all electronic states involved in the EY scattering processes. This equation represents that the spin-mixing due to SFS between spin-up and spin-down states depend on the number of spin-up ($\uparrow$) and spin-down ($\downarrow$) states at the Fermi level, which is already represented by $D_F$.

A compact differential equation regarding rate of ultrafast demagnetization dynamics as derived by Koopmans *et al.* [27], is given below:

$$\frac{dm}{dt} = Rm\frac{T_p}{T_C}(1-\coth(\frac{mT_C}{T_e}))  \tag{3}$$

where $m = M/M_S$, and $T_p$, $T_C$, and $T_e$ denote the phonon temperature, Curie temperature and electronic temperature, respectively. R is a material specific scaling factor [28], which is calculated to be:

$$R = \frac{8a_{sf}T_C^2 g_{ep}}{k_B T_D^2 D_S}, \tag{4}$$

where $a_{sf}$, $g_{ep}$, $D_S$ represent the SFS probability, coupling between electron and phonon sub-system and magnetic moment divided by the Bohr-magneton ($\mu_B$), whereas $T_D$ is the Debye temperature and $k_B$ represents the Boltzmann constant. Further, the expression for $g_{ep}$ is: $g_{ep} = \frac{3\pi D_F^2 D_P k_B T_D \lambda_{ep}^2}{2\hbar}$, where $D_P$, and $\lambda_{ep}$ denote the number of polarization states of spins and electron-phonon coupling constant, respectively, and $\hbar$ is the reduced Planck's constant. Moreover, the ultrafast demagnetization time at low fluence limit can be derived under various approximations as:

$$\tau_M = \frac{C_0 F(T/T_C)\hbar}{\pi D_F^2 \lambda_{si}^2 k_B T_C}, \tag{5}$$

where $C_0 = 1/4$, $\lambda_{si}$ is a factor scaling with impurity concentration, and $F(T/T_C)$ is a function solely dependent on ($T/T_C$) [29].

Earlier, it has been shown that a negligible $D_F$ in $CrO_2$ is responsible for large ultrafast demagnetization time. The theoretical calculation for CFMS by Oogane *et al.* shows that $D_F$ initially decreases and then increases with *x* [30] having a minima at *x* = 0.4. As $D_F$ decreases, the number of effective minority spin states become less, reducing both SOC strength, as shown by Mavropoulos et al. [31], and the effective spin-mixing parameter is given by Eq. (2), and vice versa. This will result in a reduced SFS probability and rate of demagnetization. In addition, the decrease in $D_F$ makes $g_{ep}$ weaker, which, in turn, reduces the value of R as evident from Eq. (4). As the value of R diminishes, it will slow down the rate of ultrafast demagnetization which is clear from Eq. (3). In essence, a lower value of $D_F$ indicates a lower value of R, i.e. slower demagnetization rate and larger ultrafast demagnetization time. Thus, demagnetization time is highest for $x = 0.4$. On both sides of $x = 0.4$, the value of R will increase and ultrafast demagnetization time will decline continuously. Our experimental results, supported by the existing theoretical results for the CFMS samples with varying alloy composition, clearly show that the position of Fermi level is a crucial decisive factor for the rate of ultrafast demagnetization. This happens due to the continuous tunability of $D_F$ with $x$, which causes an ensuing variation in the number of scattering channels available for SFS. To capture the effect of pump fluence on the variation of $\tau_M$, we have measured the ultrafast demagnetization curves for various applied pump fluences. All the fluence dependent ultrafast demagnetization curves are fitted with Eq. (1) and the values of corresponding $\tau_M$ are extracted. The change in $\tau_M$ with fluence is shown in Fig. 2C. A slight change in $\tau_M$ with fluence is observed which is negligible in comparison to the change of $\tau_M$ with $x$. However, this increment can be explained using the enhanced spin fluctuations at much higher elevated temperature of the spin system [28].

As the primary microscopic channel for spin angular momentum transfer is the same for both ultrafast demagnetization and magnetic damping, it is expected to find a correlation between them. We have measured the time-resolved Kerr rotation data corresponding to the magnetization precession at an applied in-plane bias magnetic field ($H_b$) of 3.5 kOe as shown in Fig. 3A. The macrospin modelling is employed to analyse the time dependent precessional data obtained by solving the Landau-Lifshitz-Gilbert equation [32] which is given below:

$$\frac{d\hat{m}}{dt} = -\gamma(\hat{m} \times H_{\text{eff}}) + \alpha(\hat{m} \times \frac{d\hat{m}}{dt}) \tag{6}$$

where $\gamma$ is the gyromagnetic ratio and is related to Lande g factor by $\gamma = g\mu_B / \hbar$. $H_{\text{eff}}$ is the total effective magnetic field consisting of $H_b$, exchange field ($H_{\text{ex}}$), dipolar field ($H_{\text{dip}}$) and anisotropy field ($H_K$). The experimental variation of precession frequency ($f$) against $H_b$ is fitted with the Kittel formula for uniform precession to extract $H_K$ values. The details of the fit are discussed in the Supplemental Materials [24].

For evaluation of $\alpha$, all the measured data representing single frequency oscillation are fitted with a general damped sine-wave equation superimposed on a bi-exponential decay function, which is given as:

$$M(t) = A + B_1 e^{-t/\tau_{fast}} + B_2 e^{-t/\tau_{slow}} + M(0) e^{-t/\tau} \sin(\omega t - \zeta) ,  \qquad (7)$$

where $\zeta$ is the initial phase of oscillation and $\tau$ is the precessional relaxation time. $\tau_{fast}$ and $\tau_{slow}$ are the fast and slow relaxation times, representing the rate of energy transfer in between different energy baths (electron, spin and lattice) following the ultrafast demagnetization and the energy transfer rate between the lattice and surrounding, respectively. A, $B_1$ and $B_2$ are constant coefficients. The value of $\alpha$ is extracted by further analysing $\tau$ using

$$\alpha = \frac{2}{[\gamma \tau (2H_b \cos(\delta - \varphi) + H_1 + H_2)]} \qquad (8)$$

where $H_1 = 4\pi M_S + \frac{2K_\perp}{M_S} - \frac{2K_1 \sin^2 \varphi}{M_S} + \frac{K_2(2 - \sin^2(2\varphi))}{M_S}$ and $H_2 = \frac{2K_1 \cos(2\varphi)}{M_S} + \frac{2K_2 \cos(4\varphi)}{M_S}$. Here $\delta$ and $\varphi$ represent the angles of $H_b$ and in-plane equilibrium $M$ with respect to the CFMS [110] axis [33]. The uniaxial, biaxial and out-of-plane magnetic anisotropies are denoted as $K_1$, $K_2$ and $K_\perp$, respectively. In our case $K_2$ has a reasonably large value while $K_1$ and $K_\perp$ are negligibly small. Plugging in all parameters including the magnetic anisotropy constant $K_2$ in Eq. (8), we have obtained the values of $\alpha$ to be 0.0041, 0.0035, 0.0046, 0.0055, 0.0061, and 0.0075 for $x = 0.00, 0.40, 0.50, 0.60, 0.75$, and 1.00, respectively. Figure 3B shows the variation of $\alpha$ with frequency for all the samples. For each sample, $\alpha$ remains constant with frequency, which rules out the presence of extrinsic mechanisms contributing to the $\alpha$. Next, we focus on the variation of $\alpha$ with $x$. Our experimental results show a non-monotonic variation of $\alpha$ with $x$ with a minima at $x = 0.4$, which is exactly opposite to the variation of $\tau_M$ with $x$. On the basis of Kambersky's SFS model [34], $\alpha$ is governed by the spin-orbit interaction and can be expressed as:

$$\alpha = \frac{\gamma \hbar^2 (\delta g)^2}{4 \Gamma M_S} D_F \qquad (9)$$

where $\delta g$ and $\Gamma^{-1}$ represent the deviation of g factor from free electron value (~2.0) and ordinary electron-phonon collision frequency. Eq. (9) suggests that $\alpha$ is directly proportional to $D_F$ and thus it becomes minimum when $D_F$ is minimum [30]. This leads to the non-monotonic variation of $\alpha$, which agrees well with earlier observation [30]. To eliminate the possible effects of $\gamma$ and $M_S$, we have plotted the variation of relaxation frequency, $G = \alpha\gamma M_S$ with $x$ which also exhibits similar variation as $\alpha$ (see the supplementary materials [24]).

Finally, to explore the correlation between $\alpha$, $\tau_M$ and alloy composition, we have plotted these quantities against $x$ as shown in Fig. 4A. We observe that $\tau_M$ and $\alpha$ varies in exactly opposite manner with $x$, having their respective maxima and minima at $x = 0.4$. Although $\tau_M$ and $\alpha$ refer to two different time scales, both of them follow the trend of variation of $D_F$ with $x$. This shows that the alloy composition-controlled Fermi level tunability and the ensuing SFS is responsible for both ultrafast demagnetization and Gilbert damping. Figure 4B represents the variation of $\tau_M$ with inverse of $\alpha$, which establishes an inversely proportional relation between them. Initially under the assumption of two different magnetic fields, i.e. exchange field and total effective magnetic field, Koopmans *et al.* theoretically proposed that Gilbert damping parameter and ultrafast demagnetization time are inversely proportional [29]. However, that raised intense debate and in 2010, Fahnle *et al.* showed that $\alpha$ can either be proportional or inversely proportional to $\tau_M$ depending upon the dominating microscopic contribution to the magnetic damping [32]. The linear relation sustains when the damping is dominated by conductivity-like contribution, whereas the resistivity-like contribution leads to an inverse relation. The basic difference between the conductivity-like and the resistivity-like contributions lies in the angular momentum transfer mechanism via electron-hole (*e-h*) pair generation. The generation of *e-h* pair in the same band, i.e. intra-band mechanism leads to the conductivity-like contribution. On the contrary, when *e-h* pair is generated in different bands (inter-band mechanism), the contribution is dominated by resistivity. Our observation of the inversely proportional relation between $\alpha$ and $\tau_M$ clearly indicates that in case of the CFMS Heusler alloy systems, the damping is dominated by resistivity-like contribution arising from inter-band *e-h* pair generation. This is in contrast to the case of Co, Fe and Ni, where the conductivity contribution dominates [35]. Typical resistivity ($\rho$) values for Co$_2$MnSi ($x = 0$) are 5 $\mu\Omega-cm$ at 5 K and 20 $\mu\Omega-cm$ at 300 K [36]. The room temperature value of $\rho$ corresponds to an order of magnitude larger contribution of the inter-band *e-h* pair generation than the intra-band generation [36]. This is in strong agreement with our experimental results and its conclusion. This firmly establishes that unlike conventional transition metal ferromagnets, damping in CFMS Heusler alloys is dominated by resistivity-like contribution, which results in an inversely proportional relation between $\alpha$ and $\tau_M$.

In summary, we have investigated the ultrafast demagnetization and magnetic Gilbert damping in the CFMS Heusler alloy systems with varying alloy composition ($x$), ranging from $x = 0$ (CMS) to $x = 1$ (CFS) and identified a strong correlation between $\tau_M$ and $x$, the latter controlling the position of Fermi level in the electronic band structure of the system. We have found that $\tau_M$ varies non-monotonically with $x$, having a maximum value of ~ 350 fs for $x = 0.4$ corresponding to the lowest $D_F$ and highest degree of spin-polarization. In-depth investigation has revealed that the ultrafast demagnetization process in CFMS is primarily governed by the composition-controlled variation in spin-flip scattering rate due to variable $D_F$. Furthermore, we have systematically investigated the precessional dynamics with variation in $x$ and extracted the value of $\alpha$ from there. Our results have led to a systematic correlation in between $\tau_M$, $\alpha$ and $x$ and we have found an inversely proportional relationship between $\tau_M$ and $\alpha$. Our thorough investigation across the alloy composition ranging from CMS to CFS have firmly established the fact that both ultrafast demagnetization and magnetic Gilbert damping in CFMS are strongly controlled by the spin density of states at Fermi level. Therefore, our study has enlightened a new path for qualitative understanding of spin-polarization from ultrafast demagnetization time as well as magnetic Gilbert damping and led a step forward for ultrafast magnetoelectronic device applications.

## Acknowledgements

This work was funded by: S. N. Bose National Centre for Basic Sciences under Projects No. SNB/AB/12-13/96 and No. SNB/AB/18-19/211.

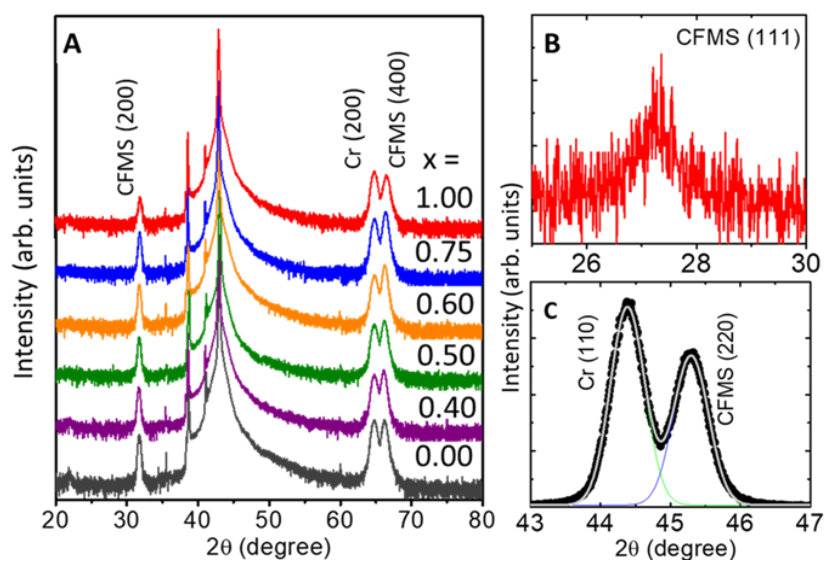

**Fig. 1.** (**A**) X-ray diffraction (XRD) patterns of $Co_2Fe_xMn_{1-x}Si$ (CFMS) thin films for different alloy composition ($x$) measured in conventional $\theta$-$2\theta$ geometry. Both CFMS (200) superlattice and CFMS (400) fundamental peaks are marked along with Cr (200) peak. (**B**) The tilted XRD patterns reveal the CFMS (111) superlattice peak for $L2_1$ structure. (**C**) CFMS (220) fundamental peak together with Cr (110) peak.

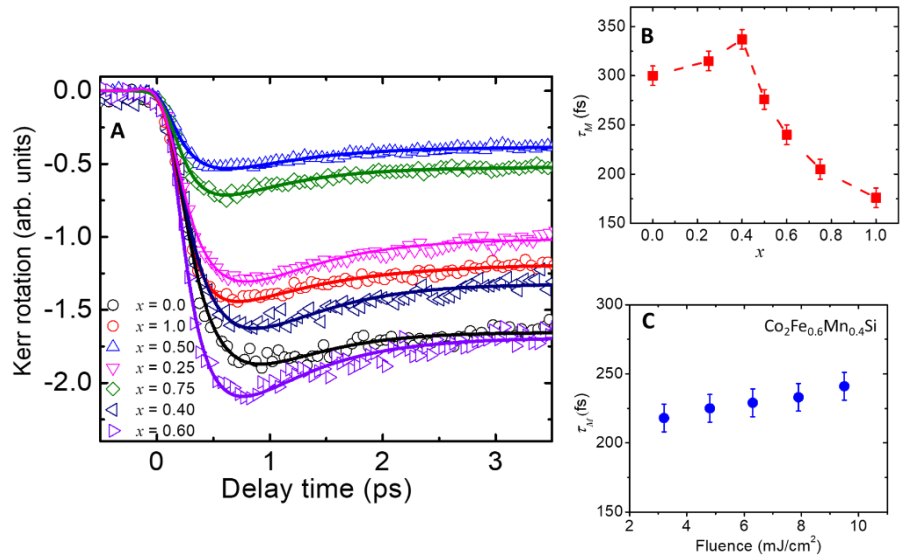

**Fig. 2.** (**A**) Ultrafast demagnetization curves for the samples with different alloy composition (*x*) measured using TRMOKE. Scattered symbols are the experimental data and solid lines are fit using Eq. 3. (**B**) Evolution of $\tau_M$ with *x* at pump fluence of 9.5 mJ/cm$^2$. Symbols are experimental results and dashed line is guide to eye. (**C**) Variation in $\tau_M$ with pump fluence.

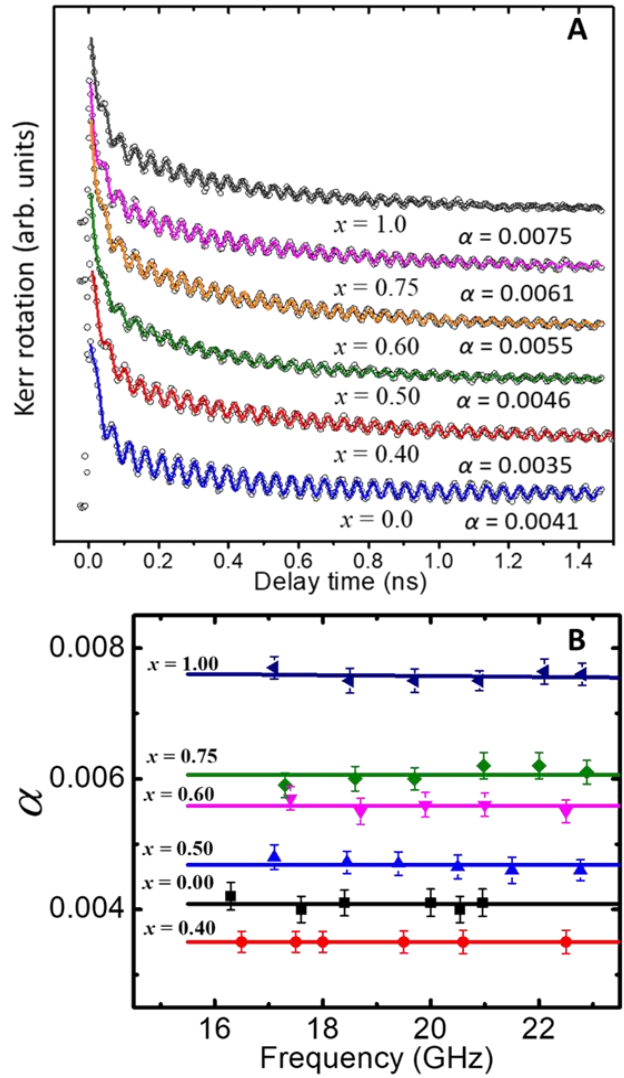

**Fig. 3.** **(A)** Time-resolved Kerr rotation data showing precessional dynamics for samples with different *x* values. Symbols are the experimental data and solid lines are fit with damped sine wave equation (Eq. 6). The extracted $\alpha$ values are given below every curve. **(B)** Variation of $\alpha$ with precession frequency (*f*) for all samples as shown by symbols, while solid lines are linear fit.

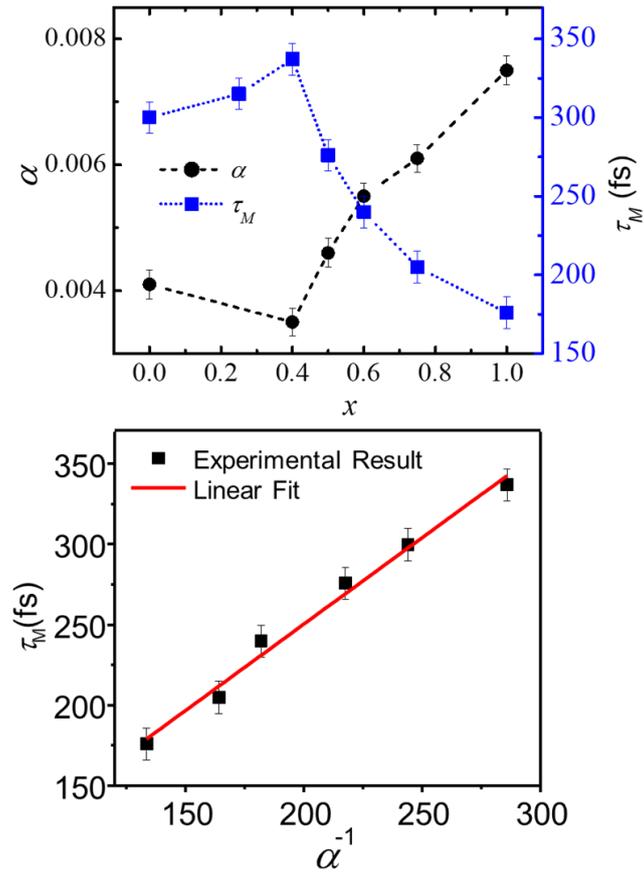

**Fig. 4. (A)** Variation of $\tau_M$ and $\alpha$ with $x$. Square and circular symbols denote the experimental results, and dashed, dotted lines are guide to eye. **(B)** Variation of $\tau_M$ with $\alpha^{-1}$. Symbols represent the experimentally obtained values and solid line refers to linear fit.

# Supplemental Materials

## I. Sample preparation method

A series of MgO Substrate /Cr (20 nm)/ Co$_2$Fe$_x$Mn$_{1-x}$Si (30 nm)/Al-O (3 nm) sample stacks were deposited using an ultrahigh vacuum magnetron co-sputtering system. First a 20-nm-thick Cr layer was deposited on top of a single crystal MgO (100) substrate at room temperature (RT) followed by annealing it at 600 ºC for 1 h. Next, a Co$_2$Fe$_x$Mn$_{1-x}$Si layer of 30 nm thickness was deposited on the Cr layer followed by an in-situ annealing process at 500 ºC for 1 h. Finally, each sample stack was capped with a 3-nm-thick Al-O protective layer. A wide range of values of $x$ is chosen, namely, $x$ = 0.00, 0.25, 0.40, 0.50, 0.60, 0.75 and 1.00. To achieve the desired composition of Fe and Mn precisely, the samples were deposited using well controlled co-sputtering of Co$_2$FeSi and Co$_2$MnSi. Direct deposition of Co$_2$Fe$_x$Mn$_{1-x}$Si on top of MgO produces strain due to lattice mismatch in the Co$_2$Fe$_x$Mn$_{1-x}$Si layer which alters its intrinsic properties [1S]. Thus, Cr was used as a buffer layer to protect the intrinsic Co$_2$Fe$_x$Mn$_{1-x}$Si layer properties [2S].

## II. Details of measurement techniques

Using *ex-situ* x-ray diffraction (XRD) measurement we investigated the crystal structure and crystalline phase of the samples. The *in-situ* reflection high-energy electron diffraction (RHEED) images were observed after the layer deposition without breaking the vacuum condition in order to investigate the epitaxial relation and surface morphology of Co$_2$Fe$_x$Mn$_{1-x}$Si layer. To quantify the values of $M_S$ and $H_C$ of the samples, we measured the magnetization vs. in-plane magnetic field (*M-H*) loops using a vibrating sample magnetometer (VSM) at room temperature with $H$ directed along the [110] direction of Co$_2$Fe$_x$Mn$_{1-x}$Si. The ultrafast magnetization dynamics for all the samples were measured by using a time-resolved magneto-optical Kerr effect (TRMOKE) magnetometer [3S]. This is a two-colour pump-probe experiment in non-collinear arrangement. The fundamental output (wavelength, $\lambda$ = 800 nm, pulse-width, $\sigma_t$ ~ 40 fs) from an amplified laser system (LIBRA, Coherent) acts as probe and its second harmonic signal ($\lambda$ = 400 nm, $\sigma_t$ ~ 50 fs) acts as pump beam. For investigating both ultrafast demagnetization within few hundreds of femtoseconds and precessional magnetization dynamics in few hundreds of picosecond time scale, we collected the time-resolved Kerr signal in two different time regimes. The time resolution during the measurements was fixed at 50 fs in -0.5 To 3.5 ps and 5 ps in -0.1 ns to 1.5 ns to trace both the phenomena precisely. The pump and probe beams were focused using suitable lenses on the sample surface with spot diameters of ~250 µm and ~100 µm, respectively. The reflected signal from the sample surface was collected and analysed using a polarized beam splitter and dual photo detector assembly to extract the Kerr rotation and reflectivity signals separately. A fixed in-plane external bias magnetic field ($H_b$) of 1 kOe was applied to saturate the magnetization for measurement of ultrafast demagnetization dynamics, while it was varied over a wide range during precessional dynamics measurement.

### III. Analysis of XRD peaks

To estimate the degree of Co atomic site ordering, one has to calculate the ratio of integrated intensity of (200) and (400) peak. Here, we fit the peaks with Lorentzian profile as shown in inset of Fig. 1S and extracted the integrated intensities as a parameter from the fit. The calculated ratio of I(200) and I(400) with respect to alloy composition (*x*) is shown in Fig. 1S. We note that there is no significant change in the I(200)/I(400) ratio. This result indicates an overall good quality atomic site ordering in the broad range of samples used in our study.

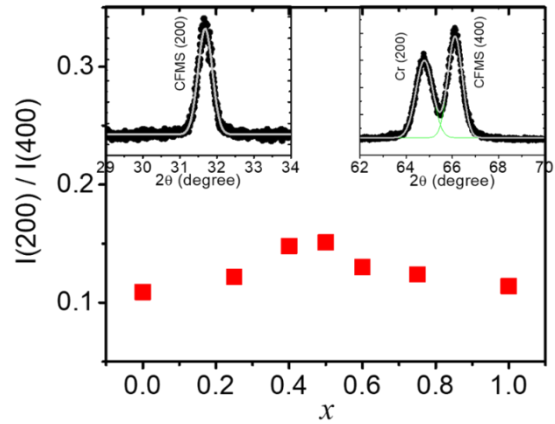

**Fig. 1S.** Variation of integrated intensity ratio I(200)/I(400) with *x*, obtained from XRD patterns. Inset shows the fit to the peaks with Lorentzian profile.

### IV. Analysis of RHEED pattern

The growth quality of the CFMS thin films was experimentally investigated using *in-situ* RHEED technique. Figure 2S shows the RHEED images captured along the MgO [100] direction for all the samples. All the images contain main thick streak lines in between the thin streak lines, which are marked by the white arrows, suggesting the formation of ordered phases. The presence of regularly-aligned streak lines confirms the epitaxial growth in all the films.

**Fig. 2S.** *In-situ* RHEED images for all the $Co_2Fe_xMn_{1-x}Si$ films taken along the MgO [100] direction. White arrows mark the presence of thin streak lines originating from the $L2_1$ ordered phase.

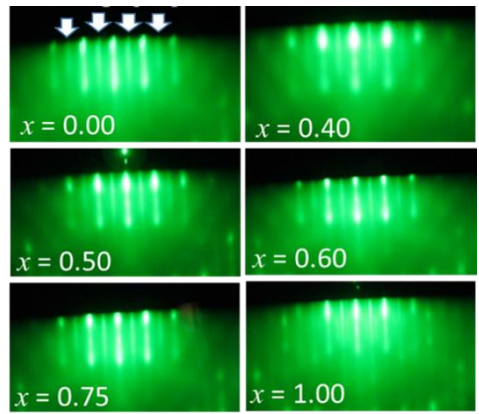

V. **Analysis of magnetic hysteresis loops**

Figure 3SA represents the *M-H* loops measured at room temperature using VSM for all the samples. All the loops are square in nature, which indicates a very small saturation magnetic field. We have estimated the values of saturation magnetization ($M_S$) and coercive field ($H_C$) from the *M-H* loops. Figure 3SB represents $M_S$ as a function of $x$ showing a nearly monotonic increasing trend, which is consistent with the Slater-Pauling rule for Heusler alloys [4S], i.e. the increment in $M_S$ due to the increase in the number of valence electrons. However, it deviates remarkably at $x = 1.0$. This deviation towards the Fe-rich region is probably due to the slight degradation in the film quality. Figure 3SC shows that $H_C$ remains almost constant with variation of $x$.

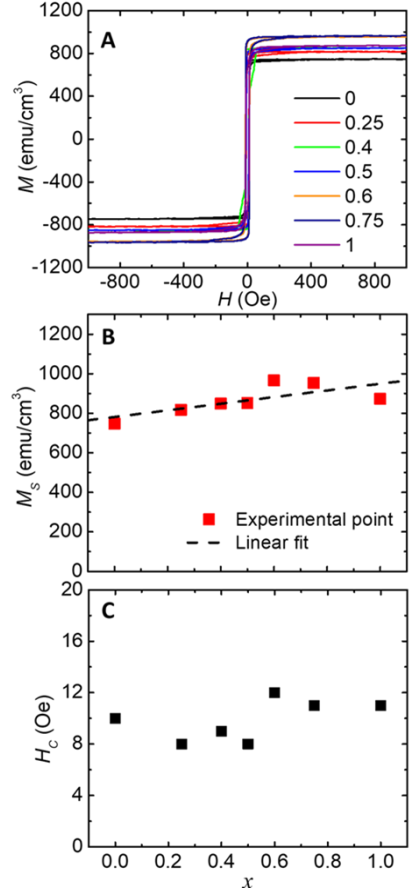

**Fig. 3S.** (**A**) Variation of *M* with *H* for all the samples. (**B**) Variation of $M_S$ as a function of *x*. Symbols are experimentally obtained values and dashed line is a linear fit. (**C**) Variation of $H_C$ with *x*.

## VI. Analysis of frequency (*f*) versus bias magnetic field ($H_b$) from TRMOKE measurements

We have experimentally investigated the precessional dynamics of all the samples using TRMOKE technique. By varying the external bias magnetic field ($H_b$), various precessional dynamics have been measured. The post-processing of these data followed by fast Fourier transform (FFT) provides the precessional frequency (*f*) and this is plotted against $H_b$ as shown in Fig. 4S.

To determine the value of in-plane magnetic anisotropy constant, obtained *f*-$H_b$ curves have been analysed with Kittel formula which is given below:

$$f = \frac{\gamma}{2\pi}\left[(H_b + 4\pi M_S + \frac{2K_2}{M_S})(H_b + \frac{2K_1}{M_S} + \frac{2K_2}{M_S})\right] \quad (1S)$$

where $M_S$ is saturation magnetization and $\gamma$ denote the gyromagnetic ratio given by $\gamma = \dfrac{g\mu_B}{\hbar}$ while $K_1$ and $K_2$ represent the two-fold uniaxial and four-fold biaxial magnetic anisotropy constant, respectively.

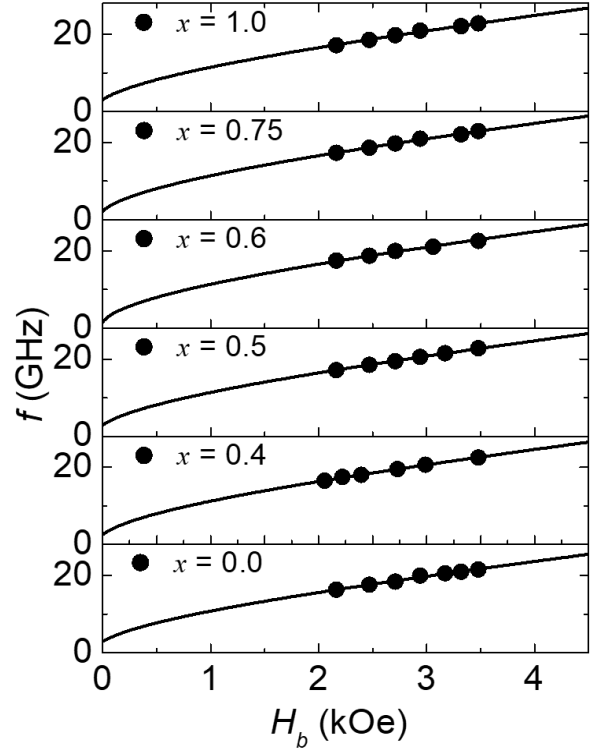

**Fig. 4S.** Variation of $f$ as a function of $H_b$. Circular filled symbols represent the experimental data and solid lines are Kittel fit.

We have found the values of several parameters from the fit including $K_1$ and $K_2$. $K_1$ has a negligible value while $K_2$ has reasonably large value in our samples. The extracted values of the parameters from the fit are tabulated as follows in Table 1S:

Table 1S: The extracted values of Lande g factor and the four-fold biaxial magnetic anisotropy constant $K_2$ for different values of $x$.

| $x$ | g | $K_2$ (erg/cm$^3$) |
|---|---|---|
| 0.00 | 2.20 | $3.1\times10^4$ |
| 0.40 | 2.20 | $2.6\times10^4$ |
| 0.50 | 2.20 | $3.0\times10^4$ |
| 0.60 | 2.20 | $2.5\times10^4$ |
| 0.75 | 2.20 | $2.6\times10^4$ |
| 1.00 | 2.20 | 3.4×10⁴ |

## VII. Variation of relaxation frequency with alloy composition

We have estimated the damping coefficient (α) and presented its variation with alloy composition ($x$) in the main manuscript. According to the Slater-Pauling rule, $M_S$ increases when the valence electron number systematically increases. As in our case the valence electron number changes with $x$, one may expect a marginal effect of $M_S$ on the estimation of damping. Thus, to rule out any such possibilities, we have calculated the variation of relaxation frequency, $G = \alpha\gamma M_S$ with $x$, which is represented in Fig. 5S. It can be clearly observed from Fig. 5S that relaxation frequency exactly follows the trend of $\alpha$. This rules out any possible spurious contribution of $M_S$ in magnetic damping.

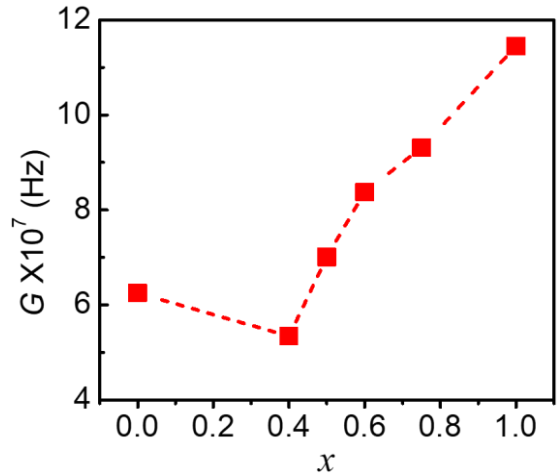

**Fig. 5S.** Non-monotonic variation of G with $x$ for all the samples.